\documentclass[aps,pra,twocolumn,nopacs,floatfix]{revtex4-2}
\usepackage{diagbox}
\usepackage{epsfig}
\usepackage{graphicx}
\usepackage{dcolumn}
\usepackage{amsthm,amsmath}
\usepackage{comment}
\usepackage[colorlinks]{hyperref}
\usepackage{xcolor}
\usepackage{longtable}
\usepackage{amsmath}

\begin{document}

\title{Electric Dipole Moments and Static Dipole Polarizabilities of Alkali--Alkaline-Earth Molecules: Non-relativistic versus relativistic coupled-cluster theory analyses}

\author{R. Mitra}
\email{ramanujmitra07@gmail.com}
\affiliation{Physical Research Laboratory, Atomic, Molecular and Optical Physics Division, Navrangpura, Ahmedabad-380009, India}
\author{V. S. Prasannaa}
\email{srinivasaprasannaa@gmail.com}
\affiliation{Centre for Quantum Engineering, Research and Education, TCG CREST, Salt Lake, Kolkata 700091, India}
\author{B. K. Sahoo}
\email{bijaya@prl.res.in}
\affiliation{Physical Research Laboratory, Atomic, Molecular and Optical Physics Division, Navrangpura, Ahmedabad-380009, India}

\begin{abstract}
We analyze the electric dipole moments (PDMs) and static electric dipole polarizabilities of the alkali--alkaline-earth (Alk-AlkE) dimers by employing finite-field coupled-cluster methods, both in the frameworks of non-relativistic and four-component spinfree relativistic theory. In order to carry out comparative analyses rigorously, we consider those Alk-AlkE molecules made out of the lightest to the  medium-heavy constituent atoms (Alk: Li to Rb and AlkE: Be through Sr). We present behaviour of electron correlation effects as well as relativistic effects with the size of the molecules. Uncertainties to the above quantities of the investigated Alk-AlkE molecules are inferred by analyzing our results from different form of Hamiltonian, basis set, and perturbative parameter in a few representative molecules. We have also provided empirical relations by connecting average polarizabilities of the Alk-AlkE molecules with their PDMs, and atomic numbers and polarizabilities of the corresponding Alk and AlKE atoms, which can be used to roughly estimate the average polarizabilities of other heavier Alk-AlkE molecules. We finally give our recommended results, and compare them with the literature values. 
\end{abstract}
\maketitle

\section{Introduction} \label{sec1}

The physics of ultracold diatomic molecules has received enormous attention over the last couple of decades for their exclusive properties~\cite{1,2,3,4} and a wide array of potential applications~\cite{5,6}. Several ultracold diatomic molecules have been successfully produced recently via different mechanisms including Feshbach resonance~\cite{Feshbach1,Feshbach2}, photoassociation~\cite{photoassociation}, deceleration of molecular beams~\cite{alloptical}, and buffer gas cooling~\cite{buffer1}. The very low temperature of ultracold molecules helps to reduce decoherence and other systematic effects, which enables one to conduct high precision measurements for investigating fundamental constants like proton to electron mass ratio~\cite{p:e1,p:e2,p:e3}, electric dipole moment of electron~\cite{eEDM1,eEDM2} et cetera. Ultracold polar molecules are also being considered as useful tools to produce qubits in the booming field of quantum computation~\cite{Demille}. 

Historically, homonuclear alkali dimers were the first set of molecules that were cooled to ultracold temperatures~\cite{alal1,alal2,alal3}. However, the focus gradually shifted from non-polar homonuclear alkali-dimers to heteronuclear alkali-dimers after the latter were successfully cooled to temperatures in the ultracold regime~\cite{uc1,uc2,uc3,uc4}. By the virtue of having permanent electric dipole moments (PDMs), the heteronuclear alkali-dimers offer the opportunity to properly interrogate and control the systems in the presence of external electric fields. Molecules possessing sufficiently large values of PDM ($\mu$) would enhance long range and anisotropic dipole-dipole interactions, which can be controlled by external electric fields~\cite{Lem}. A large $\mu$ for a given molecule also means that we require a sufficiently low electric field to align the molecular beam for understanding dipole-dipole interactions. For a given density of trapped molecules, a prior knowledge of $\mu$  would help to understand better the required dipole interaction strengths~\cite{Loh}. The electric dipole-dipole interactions have useful applications in the physics of quantum phase transitions~\cite{qp}. These dipole-dipole interactions could couple molecular electric dipoles parallel or anti-parallel to an external electric field. Each of the molecules can act as qubits and coupled molecular dipoles can be used to realize entanglements for quantum computation~\cite{Demille,Rabi,SDthesis}. Information of molecular PDM is important in understanding the chaining of molecules. The interaction strength for the chaining process of molecules in a one-dimensional optical lattice is proportional to the square of the PDM~\cite{chain}. 

Another important property that finds crucial applications in ultracold physics is the static dipole polarizability ($\alpha$) of molecules. Its magnitude plays a significant role in trapping a molecule, as the restoring force of a trapping laser beam in an optical tweezer is directly proportional to $\alpha$. Therefore, molecules with large values of $\alpha$ are more suitable for trapping and cooling~\cite{Bonin}. Moreover, for molecules trapped inside a far-off resonance optical trap, the value of $\alpha$ determines the depth of the trap depending upon the laser field intensity~\cite{Loh}. In femtosecond physics, the information on $\alpha$ plays an important role in laser-induced impulsive alignment of molecules~\cite{femto}. Hence, knowledge of both $\mu$ and $\alpha$ has significant implications in unraveling the physics of ultracold molecules.

In an earlier work~\cite{pol1}, we conducted a comparative study of PDMs and static dipole polarizabilities of alkali dimers between  non-relativistic (NR) and relativistic (Rel) theories. As alkali dimers are closed-shell systems, our next logical step is to extend the analysis to open-shell systems of interest to ultracold physics, that is, those molecules  having non-zero magnetic dipole moments (or non-zero spin). A non-zero magnetic moment would enable one to control and manipulate the systems with an external magnetic field. All the alkali (Alk) and alkaline-earth (AlkE) atoms have been successfully cooled to ultracold temperatures, thus opening up the opportunity to produce alkali--alkaline-earth (Alk-AlkE) molecules. For this reason, we chose Alk-AlkE molecules to study their $\mu$ and $\alpha$ using both NR and Rel methods. One of the first theoretical studies on Alk-AlkE molecular bond-lengths and their PDMs were carried out by Bauschlicher \textit{et al.}~\cite{Bausch}. Coupled-cluster (CC)~\cite{cc1,cc2} calculations to investigate the electronic properties of Li-AlkE molecules were performed by Kotochigova \textit{et al.}~\cite{Kotoch}. In Ref.~\cite{Abe}, the authors examined the ground state properties of eight Alk-AlkE molecules (Alk: Li, Na, K, Rb, and AlkE: Ca, Sr) using singles-doubles and partial triples CC (CCSD(T)) method, but in a NR theory framework. In another recent work \cite{Potots}, electronic and spectroscopic properties of sixteen Alk-AlkE molecules (Alk: Li, Na, K, Rb and AlkE: Be, Mg, Ca, Sr) were examined using multi-reference configuration interaction (MRCI) method. In this work, we consider Alk-AlkE molecules made from the combination of four Alk atoms (Li, Na, K, Rb) and four AlkE atoms (Be, Mg, Ca, Sr). We compare results from both the NR and Rel calculations of $\mu$ and $\alpha$ of these sixteen molecules in their ground state ($^2\Sigma^+$). This is accompanied by an analysis of correlation effects. These discussions are followed by studying the precision of our results, finding empirical relationship between PDMs and polarizabilities, and finally comparing our results with available literature values. 

This paper is organised in the following manner: in Sec.~\ref{sec2}, we expound on the theory of $\mu$ and $\alpha$, noting that we employ the  finite-field (FF) approach to calculate molecular properties in the presence of a static electric field perturbation of certain strength. In Sec. \ref{sec3}, we discuss CC theory. We also briefly describe the basis sets used in our calculations, cut-offs imposed on high-lying virtual orbitals, and other important parameters crucial to our calculations. In Sec.~\ref{sec4}, we present our results and analyses. We use atomic units (a.u.) throughout, unless stated otherwise explicitly. 

\section{Theory} \label{sec2}

The PDM (the intrinsic electric dipole moment) of a molecule can be inferred from the first-order energy shift of the ground  electronic state of the molecule in the presence of a weak static electric field. Whereas, the static dipole polarizability is defined as a property that causes a  second-order shift under the application of a weak static electric field. If the static electric field perturbation strength is $\epsilon$, the molecular ground state energy can be expressed as
\begin{eqnarray}
E_0=E_0^{(0)}+\epsilon E_0^{(1)}+\epsilon^2 E_0^{(2)}+\cdots,\label{Eq1}
\end{eqnarray}
where $E_0^{(i)}$ denotes the $i^{th}$ order energy shift, and $E_0^{(0)}$ is the unperturbed molecular ground state energy. Eq. \eqref{Eq1} can be conveniently written as
\begin{eqnarray}
E_0=E_0^{(0)}-\mu_i\epsilon_i-\frac{1}{2}\alpha_{ij}\epsilon_i\epsilon_j+\cdots,\label{Eq2}
\end{eqnarray}
where indices $i$, $j$ denote for the direction of the applied electric field perturbation and they assume integer values from 1 to 3 (with 1, 2, and 3 denoting $x$, $y$, and $z$-directions, respectively), while $\mu_i$ and $\alpha_{ij}$ are the components of PDM and second-rank static dipole polarizability tensor, respectively. Using the Taylor series expansion, Eq. \eqref{Eq1} can be rewritten as
\begin{eqnarray}
E_0=E_0^{(0)}+\epsilon_i\frac{\partial E_0}{\partial \epsilon_i}\Bigg |_{\epsilon_i\to 0}+\frac{1}{2!}\epsilon_i\epsilon_j\frac{\partial^2 E_0}{\partial\epsilon_i\partial\epsilon_j}\Bigg |_{\epsilon_i,\epsilon_j\to 0}+\cdots . \ \ \ \label{Eq3}
\end{eqnarray}
Now, comparing Eq. \eqref{Eq2} and Eq. \eqref{Eq3}, we can get the expressions for $\mu_i$ and $\alpha_{ij}$ as
\begin{eqnarray}
\mu_i=-\frac{\partial E_0}{\partial \epsilon_i}\Bigg |_{\epsilon_i\to 0}\label{Eq4}
\end{eqnarray}
and
\begin{eqnarray}
\alpha_{ij}=-\frac{\partial^2E_0}{\partial\epsilon_i\epsilon_j}\Bigg|_{\epsilon_i,\epsilon_j\to 0}.\label{Eq5}
\end{eqnarray}
Therefore, from Eq. \eqref{Eq4} and Eq. \eqref{Eq5} we see that the PDM and the components of the static dipole polarizability tensor can be expressed as first-order and second-order energy derivatives, respectively. This approach of computing energy derivatives to evaluate properties is known as the FF method. We employ the two point central difference scheme for calculating the PDMs and the corresponding three point stencil to evaluate the static dipole polarizabilities numerically. For our calculations, we choose a static electric field perturbation $\epsilon\sim 10^{-4}$ a.u.. The average static dipole polarizability ($\bar\alpha$) for a diatomic molecule is given as
\begin{eqnarray}
\bar\alpha=\frac{1}{3}(\alpha_{xx}+\alpha_{yy}+\alpha_{zz})\label{Eq6}.
\end{eqnarray}
If the chosen diatomic molecule lies on the $z$-axis, Eq. \eqref{Eq6} can be expressed as
\begin{eqnarray}
\bar\alpha=\frac{1}{3}(\alpha_{zz}+2\alpha_{xx}),\label{Eq7}
\end{eqnarray}
as here $\alpha_{xx}=\alpha_{yy}$.  $\alpha_{zz}$ is the parallel component of polarizability ($\alpha_{zz}=\alpha_\parallel$), and $\alpha_{xx}$ and $\alpha_{yy}$ constitute the perpendicular components of polarizability ($\alpha_{xx}=\alpha_{yy}=\alpha_\perp$). Therefore, Eq. \eqref{Eq7} can be expressed as
\begin{eqnarray}
\bar\alpha=\frac{1}{3}(\alpha_\parallel+2\alpha_\perp).
\end{eqnarray}
Hereafter, we shall follow this notation. Another interesting property that we report in this work, which is crucial in understanding the alignment of molecules along the applied external electric field is the polarizability anisotropy ($\Delta \alpha$) defined as the difference between the parallel and perpendicular components of polarizability
\begin{eqnarray}
\Delta\alpha=(\alpha_\parallel-\alpha_\perp).
\end{eqnarray}

\begin{table}[t]
\centering
\scriptsize
\caption{Values of PDM (in a.u.) of the considered Alk-AlkE molecules from the HF, DF, CCSD, RCCSD, CCSD(T), and RCCSD(T) methods.} \label{tab1}
\begin{tabular}{lccc}
\hline \hline
Molecule & Method & Non-relativistic & Relativistic \\
 \hline \\
 LiBe & HF/DF    & 2.04 &  2.04  \\
      & (R)CCSD  & 1.39 &  1.39  \\  
      & (R)CCSD(T) & 1.33 &   1.33  \\
 LiMg & HF/DF  &  0.33  & 0.34   \\
      & (R)CCSD & 0.36   & 0.35  \\
      & (R)CCSD(T) & 0.41   & 0.40  \\
 LiCa & HF/DF & 0.85 &  0.84  \\ 
      & (R)CCSD & 0.47 &  0.44  \\
      & (R)CCSD(T) & 0.43  & 0.40  \\
 LiSr & HF/DF & 0.30 &  0.23   \\
      & (R)CCSD & 0.18 &  0.13   \\
      & (R)CCSD(T) & 0.16  & 0.11  \\
 & & & \\
 NaBe & HF/DF & 0.09  & 0.12\\
      & (R)CCSD & 0.70 & 0.76 \\
      & (R)CCSD(T) & 0.85 & 0.86\\
 NaMg & HF/DF & 0.25  & 0.25\\
      & (R)CCSD & 0.25 & 0.24 \\
      & (R)CCSD(T) & 0.31 & 0.30\\
 NaCa & HF/DF & 0.09 &0.06 \\ 
      & (R)CCSD & 0.45 & 0.42\\
      & (R)CCSD(T) & 0.46 & 0.43\\
 NaSr & HF/DF & 0.09 &0.08 \\
      & (R)CCSD & 0.28 & 0.21\\
      & (R)CCSD(T) & 0.26 &0.20 \\
 & & & \\
  KBe & HF/DF & 0.33&0.32 \\
      &(R)CCSD & 0.58&0.57 \\
      & (R)CCSD(T) &0.76&0.75 \\
  KMg & HF/DF & 0.47 & 0.46\\
      & (R)CCSD & 0.25&0.24 \\
      & (R)CCSD(T) & 0.37&0.35 \\
  KCa & HF/DF & 0.05&0.08 \\
      & (R)CCSD & 0.66 &0.61\\
     & (R)CCSD(T) & 0.76&0.70 \\
  KSr & HF/DF & 0.16&0.13 \\
      & (R)CCSD & 0.57&0.44\\
      & (R)CCSD(T) & 0.64&0.51 \\
 & & & \\
  RbBe & HF/DF & 0.45 &0.40 \\
       & (R)CCSD & 0.49 &0.46\\
       & (R)CCSD(T) & 0.69&0.64 \\
  RbMg & HF/DF & 0.54&0.49 \\
       & (R)CCSD & 0.23 &0.21\\
       & (R)CCSD(T) & 0.37&0.33 \\
  RbCa & HF/DF & 0.09 &0.09\\
       & (R)CCSD & 0.69 &0.60\\
       & (R)CCSD(T) & 0.81 &0.70\\
  RbSr & HF/DF & 0.04&0.14 \\
       & (R)CCSD & 0.60&0.50 \\
       & (R)CCSD(T) & 0.72 &0.58 \\
\hline \hline
\end{tabular}
\end{table}

\section{Methodology} \label{sec3}

\subsection{Approximated CC methods}

As have been explained in the previous section, in order to evaluate $\mu$, $\bar\alpha$ and $\Delta\alpha$ using the FF approach, one needs to first calculate the molecular energies, which in turn require  calculating the wave function of the molecule. As the Schr\"{o}dinger (the Dirac equation in the relativistic framework) cannot be solved exactly for many-body systems, we opt for an approximate method to estimate the wave function. The starting point of our calculation is the mean-field wave function ($\arrowvert\Phi_0\rangle$), obtained by solving the Hartree-Fock (HF) equations (or the Dirac-Fock (DF) equations for the relativistic case). To rigorously include the missing electron correlation effects of the mean-field calculation in the determination of wave function and energy, we employ the CC method for both NR and Rel (RCC) framework with the reference wave function $\arrowvert\Phi_0\rangle$ chosen to be the respective HF and DF wave function. In the (R)CC method, the wave function is expressed as
\begin{eqnarray}
\arrowvert\Psi\rangle=e^T\arrowvert\Phi_0\rangle,
\end{eqnarray}
where $T=T_1+T_2+T_3+\cdots+T_{N_e}$ is the cluster operator, where $T_n$ would denote the $n$-tuple excitation operator, that is, $n$ occupied orbitals are virtually excited to $n$ unoccupied ones. $N_e$ is the total number of electrons in the molecule. In the singles and doubles excitations approximation ((R)CCSD method), we retain only $T_1$ and $T_2$ operators. In the second-quantization formalism, they are expressed as 
\begin{eqnarray}
T_1&=&\sum_{i,a}t_i^aa^\dagger_a a_i
\end{eqnarray}
and
\begin{eqnarray}
T_2&=&\frac{1}{4}\sum_{i,j,a,b}t_{i,j}^{a,b}a^\dagger_a b^\dagger_b a_j a_i,
\end{eqnarray}
where occupied orbitals are denoted by the subscripts $i,$ $j\cdots$ and virtual orbitals by $a,$ $b\cdots$, and $a_i$ defines an annihilation operator acting on the $i^{th}$ occupied orbital, while $a_a^\dagger$ refers to a creation operator acting on the $a^{th}$ virtual orbital. $t_i^a$ is the amplitude corresponding to a single excitation from $i^{th}$ occupied to $a^{th}$ virtual orbital while $t_{i,j}^{a,b}$ is the amplitude corresponding to a doubles excitation from the $i^{th}$ and $j^{th}$ occupied orbitals to $a^{th}$ and $b^{th}$ virtual orbitals, respectively. We also account for contributions from the dominant triples excitations in the perturbative approach using the (R)CCSD operators (denoted as (R)CCSD(T) method) to improve the results. 

\begin{figure*}[t]
\setlength{\tabcolsep}{2mm}
    \begin{tabular}{cc}
        \includegraphics[height=6.0cm,width=8.5cm]{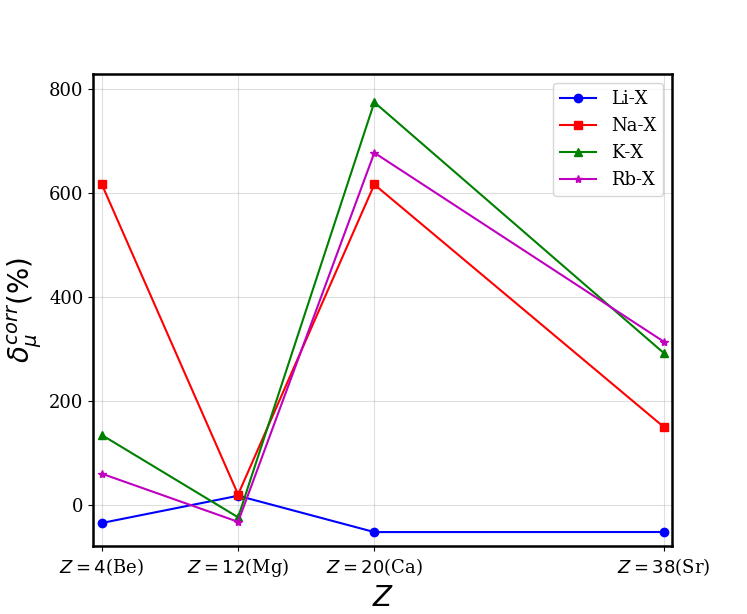} &
        \includegraphics[height=6.0cm,width=8.5cm]{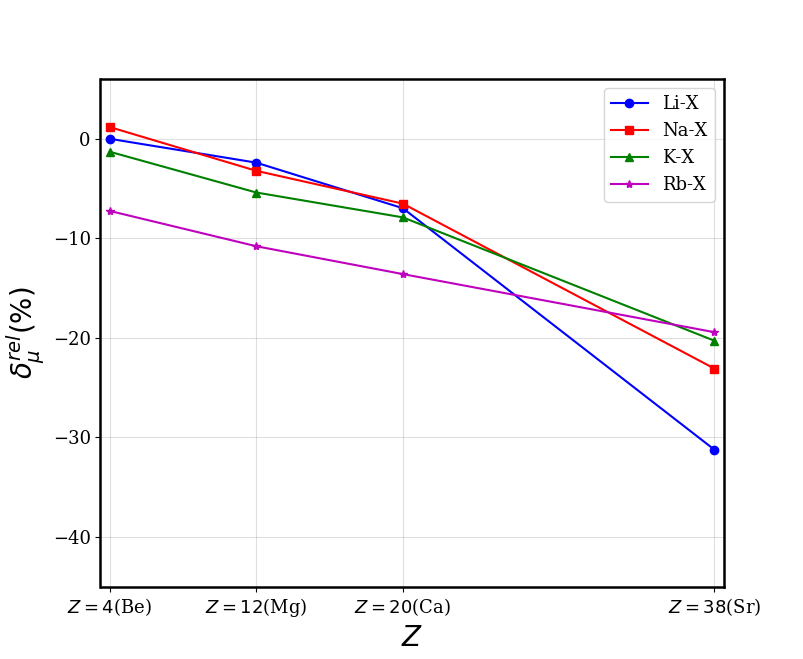} \\
        (a) &  (b)  \\
\end{tabular}
\caption{Plots demonstrating relative percentage changes in the values of $\mu$ in the Alk-AlkE molecules due to (a) the electron correlation effects ($\delta^{corr}_{\mu}$) and (b) the relativistic effects ($\delta_{\mu}^{rel}$) through the four-component spinfree Hamiltonian in the RCCSD(T) method. The x-axis shows atomic number of the alkaline-earth atoms. }
\label{fig1}
\end{figure*}

\begin{table*}[t]
\centering
\scriptsize
    \caption{Values of different components of the electric dipole polarizabilities ($\alpha_{\parallel}$ and $\alpha_{\perp}$) as well as average polarizability ($\bar \alpha$) and polarizability anisotropy ($\Delta \alpha$) of the considered Alk-AlkE molecules from the HF, DF, CCSD, RCCSD, CCSD(T), and RCCSD(T) methods. All units are in a.u..} 
    \label{tab2}
\begin{tabular}{p{1.3cm}p{2.4cm}p{1.3cm} p{1.3cm} p{1.3cm} p{1.3cm} p{0.2cm} p{1.3cm}p{1.3cm}p{1.3cm}p{1.3cm}p{1.3cm}p{1.3cm}} 
\hline \hline
 & & \multicolumn{4}{c}{Non-relativistic} & & \multicolumn{4}{c}{Relativistic} \\
\cline{3-6} \cline{8-11} \\
Molecule & Method & $\alpha_\parallel$&$\alpha_\perp$&$\bar\alpha$&$\Delta\alpha$ & & $\alpha_\parallel$&$\alpha_\perp$&$\bar\alpha$&$\Delta\alpha$ \\
 \hline \\
 LiBe&HF/DF& 211.36&88.52&129.46&122.84 & &212.06 &88.56 &129.73&123.5\\
 &(R)CCSD &373.58&111.54&198.87&262.04& & 374.13&111.6&199.11&262.53\\
 &(R)CCSD(T) &376.55&114.42&201.78 &  262.13& & 376.9&114.12&201.91&262.48\\
 LiMg& HF/DF &557.42&197.55&317.51&  359.87& & 553.14&197.66&316.15&355.48\\
 & (R)CCSD &497.42&169.19&278.6&328.23 & &495.95&169.37&278.23&326.58\\
 & (R)CCSD(T) &481.64&166.64&271.64&315 & &480.63&166.84&271.44&313.79\\
  LiCa& HF/DF &445.36&252.82&317&192.54 & &455.26&250.32&318.63&204.94\\
  &(R)CCSD &559.03&231.68&340.8&327.35& & 563.01&229.97&340.98&333.04\\
  &(R)CCSD(T) &580.16&229.88&346.64&350.28 & &584.28&228.08&346.81&356.2\\
  LiSr&HF/DF &504.27&332.4&389.69&171.87& & 464.91&296.6&352.7&168.31\\
  &(R)CCSD &597.14&283.13&387.8&314.01& & 560.36&268.63&365.78&291.73\\
  &(R)CCSD(T) &622.61&276.08&391.59&346.53& & 620.03&268.5&385.68&351.53\\
 & & & & & \\
  NaBe&HF/DF &532.87&153.96&280.26&378.91& & 536.96&147.47&277.3&389.49\\
  &(R)CCSD &396.7&144.08&228.27&252.62& & 402.3&140.48&227.75&261.82\\
  &(R)CCSD(T) &390.48&137.79&222.02&252.69& & 392.99&140.2&224.46&252.79\\
  NaMg&HF/DF &490.11&220.49&310.36&269.62& & 485.25&219.52&308.1&265.73\\
  &(R)CCSD &446.77&186.06&272.96&260.71& & 442.72&185.37&271.15&257.35\\
  &(R)CCSD(T) &441.11&183.04&269.06&258.07& & 437.29&182.39&267.36&254.9\\
  NaCa&HF/DF &743.08&276.5&432.03&466.58& & 730.78&274.73&426.75&456.05\\
  &(R)CCSD &605.78&246.26&366.1&359.52& & 600.19&245.06&363.44&355.13\\
  &(R)CCSD(T) &590.94&243.6&359.38&347.34& & 585.93&242.25&356.81&343.68\\
  NaSr&HF/DF &816.5&325.57&489.21&490.93& & 784.82&316.82&472.82&468\\
  &(R)CCSD &666.07&299.65&421.79&366.42& & 648.63&284.99&406.2&363.64\\
  &(R)CCSD(T) &652.97&296.04&415.02&356.93& & 636.1&280.9&399.3&355.2\\
  & & & & & \\
  KBe&HF/DF &763.69&433.35&543.46&330.34& & 745.42&424.52&531.49&329.9\\
  &(R)CCSD &646.62&276.35&399.77&370.27& & 628.81&271.91&390.88&356.9\\
  &(R)CCSD(T) &638.91&250.30&379.84&388.61& & 621.5&246.63&371.59&374.87\\
  KMg&HF/DF &764.42&451.98&555.86&312.84& & 749.44&443.96&545.79&305.48\\
  &(R)CCSD &671.65&315.14&433.98&356.51& & 657.19&310.20&425.86&346.99\\
  &(R)CCSD(T) &658.97&295.79&416.85&363.18& & 644.47&291.31&409.03&353.16\\
  KCa&HF/DF &1209.95&462.82&711.86&747.13& & 1171.9&468.66&703.07&703.24\\
  &(R)CCSD &956.7&347.2&550.37&609.5& & 931.75&350.76&544.42&580.99\\
  &(R)CCSD(T) &909.43&330.71&523.62&578.72& & 888.38&334.4&519.06&553.98\\
  KSr&HF/DF &1339.52&505.45&783.47&834.07& & 1249.65&498.21&748.69&751.44\\
  &(R)CCSD &1026.44&388.06&600.85&638.38& & 975.42&387.44&583.43&587.98\\
  &(R)CCSD(T) &971.74&372.68&572.37&599.06 & & 928.23&372.13&557.5&556.1\\
    & & & & & \\
  RbBe&HF/DF &819.56&554.86&643.09&264.7& & 773.14&507.91&576.32&265.23\\
  &(R)CCSD & 697.67&333.56&454.93&364.11& & 648.61&312.05&424.24&336.56\\
  &(R)CCSD(T)&692.69&295.05&427.6&397.64& & 644.08&278.14&400.12&365.94\\
  RbMg&HF/DF &862.3&567.18&665.55&295.12& & 806.75&523.44&617.88&283.31\\
  &(R)CCSD& 740.42&366.43&491.09&373.99& & 684.47&343.39&457.08&341.08\\
  &(R)CCSD(T) &720.48&335.83&464.05&384.65& & 667.39&316.44&433.42&350.95\\
  RbCa&HF/DF &1372.88&567.82&836.17&805.06& & 1260.82&533.44&775.9&727.38\\
  &(R)CCSD &1072.62&393.68&619.99&678.94& & 986.42&380.38&582.39&606.04\\
  &(R)CCSD(T) &1015.59&368.26&584.04&647.33& & 939.92&357.71&551.78&582.21\\
  RbSr&HF/DF &1526.59&599.47&908.51&927.12& & 1349.35&562.62&824.86&786.73\\
  &(R)CCSD &1150.27&438.2&675.56&712.07& & 1027.13&407.66&614.25&619.77\\
  &(R)CCSD(T) &1080.95&414.26&636.49&666.69& & 970.25&385.59&580.48&584.66\\
\hline \hline
    \end{tabular}
\end{table*}

\subsection{Bond-lengths and basis functions}

We have used the equilibrium bond-lengths of the Alk-AlkE molecules from Ref. ~\cite{Potots}, where the authors carried out  MRCI calculations to obtain the ground state energies of the chosen molecules, and obtained the equilibrium bond-lengths from the minima of the corresponding potential energy curves. For lighter elements (Li, Be, Na, Mg), we used correlation-consistent polarized core-valance quadruple zeta (cc-pCVQZ) basis sets~\cite{ccpcvqz1,ccpcvqz2}, and for the heavier ones (K, Ca, Rb, Sr), we used Dyall's quadruple zeta (QZ)~\cite{DBS} basis functions. For the reduction of computational cost, at the (R)CC level, we have cut-off virtual orbitals having energies greater than 1000 a.u. for the relatively heavier molecules (NaSr, KSr, RbBe, RbMg, RbCa and RbSr). All the computations were carried out using the DIRAC program~\cite{dirac}. 

\section{Results and discussion} \label{sec4}

We investigate the trends in relativistic as well as electron correlation effects to the undertaken properties in the Alk-atom family-wise. For example, Li-family refers to LiX molecules, where X could be Be, Mg, Ca, or Sr atom. The rationale behind categorizing the chosen molecules as families rather than looking for trends based on electron number is due to the fact that two molecules that are next to each other in the number of electrons can display very dissimilar trends, since one molecule may contain a combination of light-heavy nuclei, while the other moderate-moderate nuclei. There is no particularly strong reason to provide our results based on Alk-atom family rather than AlkE-atom family, and therefore when certain trends are easier to see with the latter, we have appropriately discussed them in the main text. For the purpose of analyzing the trends, we use two quantities in the subsequent sub-sections -- one for correlation effects and the other for relativistic ones. For the former, we define  $\delta^{corr}_{P}= \left ( \frac{P_{(R)CC}-P_{DF/HF}}{P_{DF/HF}}\times 100 \right )$ that signifies the relative percentage of correlation contributions to the property, $P$. For the latter, we define  $\delta^{rel}_P=\left (\frac{P_{Rel}-P_{NR}}{P_{NR}}\times 100 \right )$. 

\subsection{Results for PDM}

In Table \ref{tab1}, we present the values of PDM for all the considered systems from both the NR and Rel calculations. To demonstrate the roles of electron correlation effects explicitly, we give values from the HF/DF, (R)CCSD, and (R)CCSD(T) methods for all the molecules. We observe from the table that electron correlation plays a significant role in determining the quality of the final values. These effects are largest in the RCCSD(T) values for NaCa (where the percentage fraction correlation is about 620), KCa (775), and RbCa (680). In the LiX family, the magnitudes of the PDM  decrease gradually from the HF method to the CCSD(T) method, except in the case of LiMg, which shows the opposite behaviour. However, this trend changes for NaX family, where the values increase from the HF method to the CCSD(T) method, with two exceptions -- NaSr, where the PDM increases from HF to CCSD, but decreases slightly (within error margins) in the CCSD(T) method, and NaMg, where the relativistic results decrease ever so slightly (again within error margins) from HF to CCSD, and then show a clear increase in the magnitude of the PDM. For both the KX and RbX families, the values increase from the HF method to the CCSD(T) method gradually, except for KMg and RbMg. In summary, for non-Mg containing molecules, we observe from our results for PDM that except for the lightest Li family, where correlation effects decrease the PDM, the general tendency of electron correlation is to increase the value of the property. For those chosen molecules that contain Mg, correlation effects increase the PDM in the lighter Li and Na families, whereas for the heavier K and Rb families, correlation decreases the PDM. Fig.~\ref{fig1}(a), which plots the related quantity, $\delta^{corr}_\mu$ for the relativistic case for the four families, shows these observed trends. An interesting observation that arises from the distribution of $\delta^{corr}_{\mu}$ for the Mg family in the figure is that $|\mu_{corr}/\mu_{HF}|$ is almost a constant, especially relative to the same quantity for the other families. 

 \begin{figure*}
   \setlength{\tabcolsep}{2mm}
     \begin{tabular}{cc}
    \includegraphics[height=6.0cm,width=8.5cm]{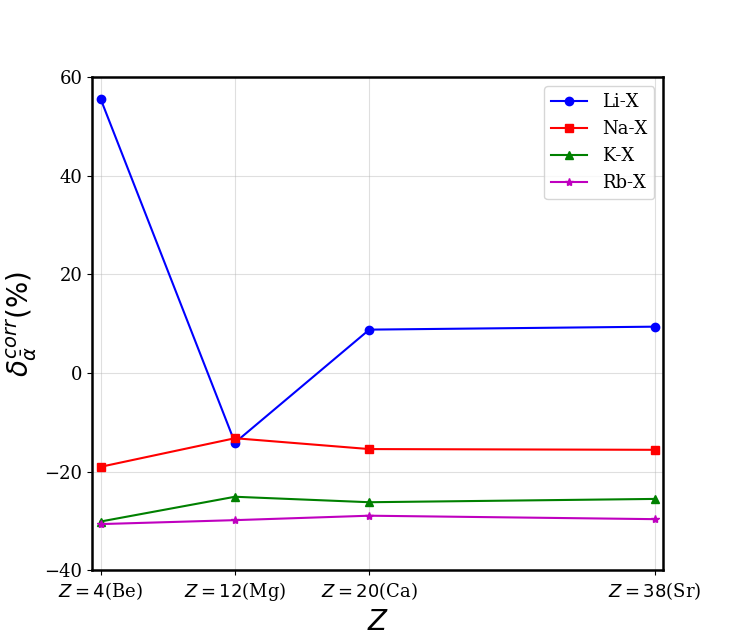}
    & \includegraphics[height=6.0cm,width=8.5cm]{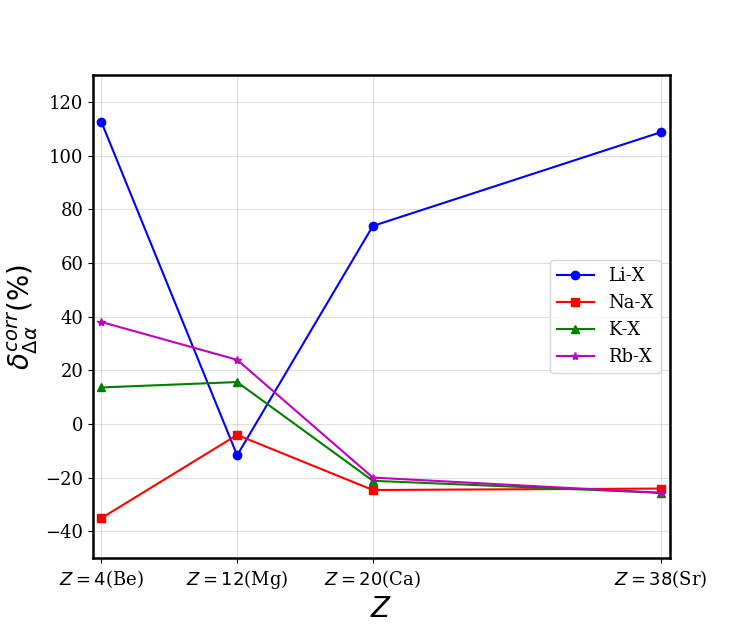} \\
         (a) &  (b)  \\
    \end{tabular}
    \caption{Schematic figures showing relative percentage changes in (a) the ${\bar{\alpha}}$ values and (b) the ${\Delta{\alpha}}$ values of the Alk-AlkE molecules due to the electron correlation effects at the RCCSD(T) method. The values in y-axis are given in \% while x-axis shows atomic number of the alkaline-earth atom for a given alkali atom family.}
    \label{fig2}
\end{figure*}

\begin{figure*}
    \setlength{\tabcolsep}{2mm}
        \begin{tabular}{cc}
    \includegraphics[height=6.0cm,width=8.5cm]{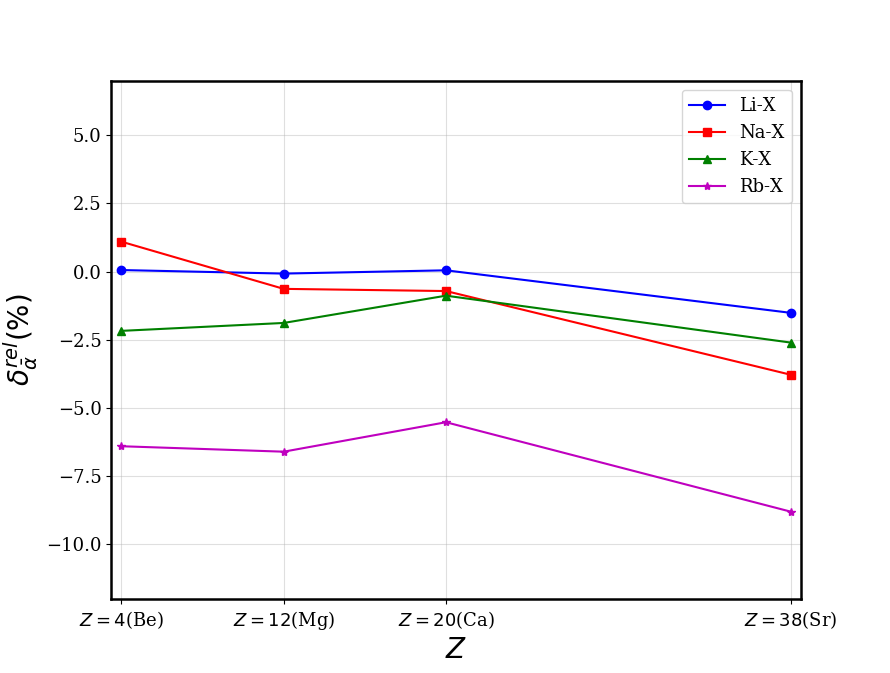} &  \includegraphics[height=6.0cm,width=8.5cm]{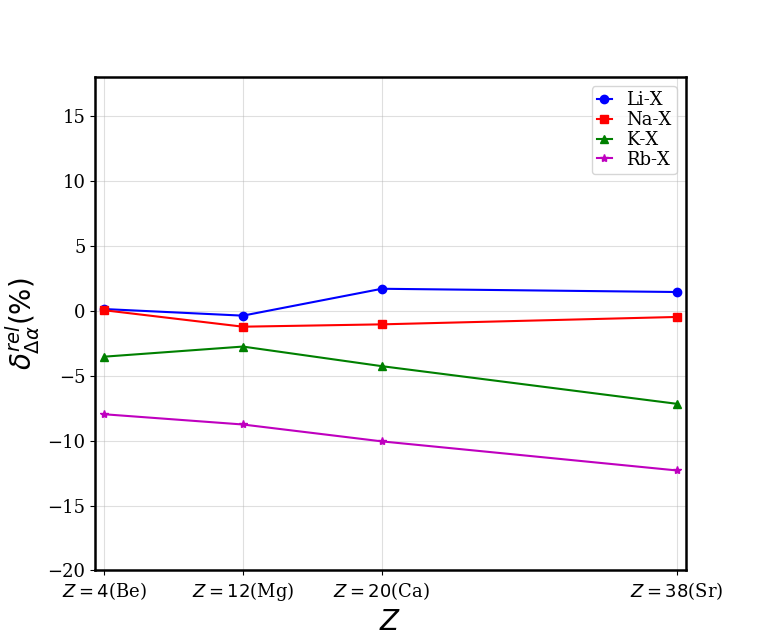}\\
        (a) &  (b)  \\
        \end{tabular}
    \caption{Schematic figures showing relative percentage changes in (a) ${\delta_{\bar{\alpha}}^{rel}}$, and (b) ${\delta_{\Delta{\alpha}}^{rel}}$ in the Alk-AlkE molecules using the RCCSD(T) method. The x-axis provides atomic number of the alkaline-earth atom, as in the previous figure.}
\label{fig3}
\end{figure*}

Having discussed correlation effects, we now compare the results for PDM between the non-relativistic and relativistic methods, from the data presented in Table~\ref{tab1} and Fig.~\ref{fig1}(b), with the figure plotting percentage fraction difference due to relativity at the CCSD(T) level of theory. The first observation that we can draw from our results is that the effect of relativity is to lower the value of PDM in most cases. The only exceptions are the lightest LiBe, where there is practically no difference between NR and Rel values, and NaBe, where relativistic effects are around 1 percent. This is reflected in Fig.~\ref{fig1}(b), where for each of the families, the slope is negative. The figure also shows that in each family, the importance of relativity increases with the atomic number of the AlkE atom. However, the rate of change is non-trivial, with crossings observed between families. We also note that the importance of relativistic effects is most pronounced in the heaviest molecules considered of each family, with the PDM of LiSr changing the most with the inclusion of relativistic effects (31 percent). In summary, while the PDM is lowered when one switches to a relativistic framework from a non-relativistic one, the degree to which the quantity decreases increases with the atomic number of a given AlkE atom. Note that this is not directly obvious visually from Fig.~\ref{fig1}(b), since the x-axis is not linear. 

Relativistic effects impact the results both at mean field and the correlation levels of theory. In the former, we see that relativity accounts for as much as 250 percent for RbSr, but also note that this is an exception; they are at most about 33 percent in the the rest of the molecules. In the correlation sector, inclusion of relativity can decrease the correlation value by as large as 35 percent for RbSr. 

\subsection{Results for polarizability}

In Table~\ref{tab2}, we present values of $\alpha_\parallel$, $\alpha_\perp$, $\bar\alpha$, and $\Delta\alpha$, both from NR and Rel calculations, and in each of these cases, we provide results at mean field, CCSD, and CCSD(T) levels of theory. We note that the individual values as well as the trends for $\bar{\alpha}$ and $\Delta \alpha$ are obviously influenced by those in $\alpha_{\parallel}$ and $\alpha_{\perp}$. Therefore, in view of the former two quantities being the ones that are obtained in experiment, we focus on the individual values and trends of these properties. Otherwise, the discussion of the flow of results in this sub-section will be very similar to that from the previous one. 

We begin with a discussion on correlation effects, based on data from Table~\ref{tab2} and Fig.~\ref{fig2}. We expect the trends in polarizabilities to be at least somewhat different from those in PDMs not only because the quantities are intrinsically different, but also because the stress would not be on $\alpha_\parallel$ and $\alpha_\perp$, but rather on $\bar\alpha$ and $\Delta\alpha$. We observe that electron correlation influences the results significantly, with the largest percentage fraction difference between the RCCSD(T) and DF results occurring for the lightest LiBe molecule (about 60 percent for $\bar \alpha$ and about 110 percent for $\Delta \alpha$). However, it is not as striking as in the case of PDMs, where we encountered changes as large as about 775 percent. Trend-wise, the plots in sub-figures 2(a) and 2(b) display an almost monotonic trend for each of the families, with the molecules containing Be in each family being an exception, unlike in the case of PDMs. Note that in the non-relativistic case, the correlation trends are similar for $\Delta \alpha$, but are slightly different in the case of $\bar{\alpha}$. One can also observe from sub-figures 2(a) and 2(b) that with the exception of the Li family, the correlation trends vary relatively mildly for the rest of the families for $\bar{\alpha}$, whereas for $\Delta \alpha$, we see more variation. This indicates that accurate determination of the values of $\Delta\alpha$ for the Alk-AlkE systems would be more sensitive to proper inclusion of the electron correlation effects. We observe similar signs of the peculiarity seen in the Mg family of Fig.~\ref{fig1} in our relativistic results, and in this case, we find that $|{\Delta \alpha}_{corr}/{\Delta \alpha}_{HF}|$ varies within around 30 percent, which is not small but still significantly smaller relative to other families. The same quantity constructed for $\bar \alpha$ also varies within 30 percent across the Mg family, which is again much lesser than the maximum variation seen in each of the other families. In summary, we find that correlation effects play a very important role in determining $\bar{\alpha}$ and $\Delta \alpha$, but not as extreme as in the case of PDMs. We also observe that the trends for both $\bar{\alpha}$ and $\Delta \alpha$ are smoother than the trend found for the PDMs. Lastly, we see a similar behaviour in $|P_{corr}/P_{HF}|$ for the Mg family as it was for PDMs. 

\begin{table}[t]
    \centering
     \caption{The values of PDM and parallel component of polarizabilities of the LiBe, KBe, and RbBe molecules using different Hamiltonians in the RCCSD(T) method. All units are in a.u.. }
    \begin{tabular}{ccc}
    \hline\hline \\
      Molecule  &$\mu$ &$\alpha_{\parallel}$ \\
        \hline \\
        &\multicolumn{2}{c}{Dirac-Coulomb}\\
        \cline{2-3}\\
        LiBe&1.33&376.96\\
        KBe&0.75&621.64\\
        RbBe&0.64&644.01\\
        \hline\\        
        &\multicolumn{2}{c}{4-component spinfree}\\
        \cline{2-3}\\
        LiBe&1.33&376.9\\
        KBe&0.75&621.5\\
        RbBe&0.64&644.08\\
        \hline
        &\multicolumn{2}{c}{Spinfree-X2C}\\
        \cline{2-3}\\
        LiBe&1.32&376.86\\
        KBe&0.75&621.76\\
        RbBe&0.64&644.58\\
        \hline \\
        &\multicolumn{2}{c}{DKH2}\\
        \cline{2-3}\\
        LiBe&1.32&376.99\\
        KBe&0.72&620.31\\
        RbBe&0.64&644.87\\
        \hline\\
        &\multicolumn{2}{c}{Non-relativistic}\\
        \cline{2-3}\\
        LiBe&1.32&376.55\\
        KBe&0.76&638.91\\
        RbBe&0.69&692.69\\
        \hline\hline\\
    \end{tabular}
    \label{tab3}
\end{table} 

We will now turn our attention to the importance of relativistic effects on polarizabilities. From Fig.~\ref{fig3}(a), we immediately notice that the effect of relativity is to decrease the value of $\bar{\alpha}$ in each of the families, with the Li family being an exception. On the other hand, relativity increases the value of $\Delta \alpha$ for six out of the sixteen systems considered. The trends are a reflection of those in $\alpha_{\parallel}$ and $\alpha_{\perp}$, as $\bar{\alpha}$ involves a sum of these quantities, while $\Delta \alpha$ is constructed from the difference between them. When we compare the importance of relativistic and correlation effects, we find that while the former can be as large as 10 percent and therefore not negligible in their own right, the latter is more pronounced, and is as large as 120 percent for LiBe. In summary, the relativistic trends in $\bar{\alpha}$ and $\Delta \alpha$ are not as clear as in the case of PDM, and relativistic effects are not as dominant as correlation effects, while being non-negligible on their own. 

\begin{table}[t]
    \centering
    \caption{The values of PDM and polarizability of four representative molecules (LiBe, KBe, RbBe, and RbSr) with basis sets of increasing cardinal number (DZ to QZ) in the RCCSD(T) method, and with the four-component spinfree relativistic Hamiltonian. We also give results from a CBS extrapolation scheme. All units are in a.u..}
    \begin{tabular}{ccccccc}
    \hline\hline
         Molecule&Basis &$\mu$ &$\alpha_\parallel$ &$\alpha_\perp$ &$\bar\alpha$&$\Delta\alpha$ \\
         \hline \\
         LiBe&DZ&1.1&387.33&114.73&205.6&272.6\\
         &TZ&1.27&378.12&114.06&202.08&264.06\\
         &QZ&1.33&376.9&114.12&201.91&262.48\\
         &CBS&1.37&376.01&114.69&201.8&261.32 \\
  & & & \\
         KBe&DZ&0.35&555.73&257.82&357.12&297.91\\
         &TZ&0.65&608.5&248.04&368.19&360.46\\
         &QZ&0.75&621.5&246.63&371.59&374.87\\
         &CBS&0.81&630.99&245.6&374.06&385.39\\
  & & & \\
        RbBe&DZ&0.27&577.9&289.86&385.87&288.04\\
        &TZ&0.55&627.78&279.22&395.41&348.56\\
        &QZ&0.64&644.08&278.14&400.12&365.94\\
        &CBS&0.71&655.97&277.34&403.55&378.63\\
 & & & \\
         RbSr&DZ&0.26&915.82&387.95&563.91&527.87\\
         &TZ&0.52&957.48&384.35&575.39&573.13\\
         &QZ&0.58&970.25&385.59&580.48&584.66\\
         &CBS&0.62&979.56&386.49&584.18&593.07\\
         \hline\hline
    \end{tabular}
    \label{tab4}
\end{table}


\subsection{Reliability tests}

In order to present our recommended values  along with their uncertainties from our calculations, we first assess the reliability of our results in the form of multiple precision checks. The precision of a calculation would depend on the choice of Hamiltonian and the wave function. The choice of wave function entails selecting the single particle basis as well as the quantum many-body theory to employ. Further, since we adopt the FF approach, we also need to check the dependence of the precision in our results on the choice of perturbing parameter. We will address each of these aspects very briefly in the next few paragraphs, in order to arrive at a reasonable estimate for the recommended values of the properties for the chosen molecules. Note that choosing a better stencil than our simplest central difference  scheme would have little effect on results given that the perturbing parameter, $\epsilon$, is already small. For carrying out these reliability tests, we choose three representative systems, which include the lightest LiBe, the slightly heavier KBe, and the moderately heavy RbBe. In view of the steep computational cost, we did not choose the heavier systems for reliability tests. 

The first of our precision checks is the influence of the choice of Hamiltonian. For this analysis, we assume that our RCCSD(T) values are the more accurate results compared to those from the RCCSD method due to the fact that the former takes into account more physical effects. We consider a hierarchy of Hamiltonians in terms of physical effects included as well as the resources that they consume. They include the computationally very expensive four-component Dirac-Coulomb Hamiltonian, the four-component spinfree Hamiltonian that we used for all our main results, an exact two-component Hamiltonian (spinfree X2C)~\cite{Tsaue}, an approximate two-component Hamiltonian (we choose second-order Douglas-Kroll-Hess \cite{DK,Hess,Reiher} Hamiltonian for this purpose), and finally the non-relativistic Hamiltonian. Note that the four-component Dirac-Coulomb Hamiltonian (DCH) does not include the Coulomb integrals of the type SSSS (where S stands for small component), but rather replaced by interatomic corrections~\cite{Visscher97,Viss1997a}. This approximation is known to be very accurate, and save lots of computational time. In spite of this approximation, the DCH is very expensive. In view of this situation, we only test the dependence of results on the choice of Hamiltonian for $\mu$ and $\alpha_{\parallel}$, and assume that the difference in results between any two chosen Hamiltonians would be comparable for $\alpha_{\perp}$ too. The results in Table~\ref{tab3} show that the PDM of the considered systems are practically unchanged between the DCH and our four-component spinfree calculations. We also observe a similar behaviour for $\alpha_{\parallel}$. This indicates that the effect of spin-orbit coupling is negligible in these systems, and we expect a similar behaviour to hold for the other Alk-AlkE molecules. We also see that the results agree very well with the spinfree X2C Hamiltonian. Note that spinfree X2C Hamiltonian, which is a two-component Hamiltonian, still contains in it relativistic effects (excluding spin-orbit coupling). We also see no noteworthy deviations when we compare our results with DKH2 Hamiltonian. We take this a step further by carrying out calculations of the PDM and $\alpha_\parallel$ for the heaviest RbSr molecule with spinfree X2C Hamiltonian, and find that the PDM remains unchanged while $\alpha_\parallel$ changes well within 1 percent. 

\begin{table}[t]
    \centering
    \caption{The PDM and dipole polarizabilities (all in atomic units) of the LiBe, KBe, and RbBe molecules with different choices of the perturbing parameter, $\epsilon$.}
    \begin{tabular}{ccccc}
    \hline\hline
         Molecule&$\epsilon$&$\mu$ &$\bar\alpha$&$\Delta\alpha$\\
        \hline \\
        LiBe&0.00005&1.33&203.77&259.68\\
         &0.0001&1.33&201.91&262.48\\
         &0.0005&1.32&201.29&263.92\\
  & & & \\
         KBe&0.00005&0.75&370.42&376.64\\
         &0.0001&0.75&371.59&374.87\\
         &0.0005&0.75&372.08&372.98\\
  & & & \\
         RbBe&0.00005&0.64&404.22&359.91\\
         &0.0001&0.64&400.12&365.94\\
         &0.0005&0.64&399.18&366.73\\
         \hline\hline
    \end{tabular}
    \label{tab5}
\end{table}

Another important test of reliability of our calculations is to analyze the basis set dependence of the results. To that end, we have performed calculations using the four-component spinfree relativistic Hamiltonian in the RCCSD(T) method with Dyall's triple-zeta (TZ) and double-zeta (DZ) basis functions. Again, we have considered LiBe, KBe and RbBe as representative molecules. Further, we carry out a two-point complete basis set (CBS) \cite{Helg,Gome} extrapolation with our TZ and QZ results upto $\zeta=50$. We enlist the RCCSD(T) values of $\mu$, $\bar\alpha$ and $\Delta\alpha$ using the aforementioned basis sets in Table~\ref{tab4}. The table shows that $\mu$ increases and converges systematically from cardinal numbers 1 through 50. However, we see a relatively strong dependence of PDMs on basis, as the results change from QZ to the CBS limit by about  $4\%$ for LiBe, 8$\%$ for KBe, and 11$\%$ for RbBe. We extended our analysis to the heaviest RbSr molecule in view of this dependence being important, and found that the percentage fraction difference between results using QZ basis and with CBS extrapolation is  7.5$\%$. We carry out the same analysis for polarizabilities too, and we find that the QZ basis is reasonably good, with the percentage fraction differences between QZ and CBS extrapolation being utmost $1\%$ for $\bar{\alpha}$ and $4\%$ for $\Delta \alpha$. 

The other possible source of uncertainty in our calculation could stem from the contributions from high-lying orbitals that are not considered to account for electron correlation effects in the RCCSD(T) method. To carry out the calculations with the available computational resources, we have imposed a cut-off to virtual orbitals having more than energies at 1000 a.u. for relatively heavier systems. To get an estimate of error arising from the neglected higher virtual orbitals, we calculate PDMs and polarizabilities for RbBe,  by considering all the generated virtual orbitals in the QZ basis. We found that the changes in the values are insignificant and within the precision of the present interest. This further assured us about the precision in the calculations. 

We now consider the error due to the missing  triple excitations in the CC method. Since CCSD(T) is widely regarded as a gold standard for molecular property calculations, we expect that the error due to missing triple excitations from RCCSD(T) would be much lesser than the percentage fraction difference between the RCCSD and the RCCSD(T) values, and assume that it would be roughly half its magnitude or lower. We find that the percentage fraction difference between the RCCSD and the RCCSD(T) methods is at most about 6 percent for $\bar{\alpha}$, and is within 7 percent for $\Delta \alpha$, except in the case of LiSr, where it is significant at about 17 percent. However, for the PDM, we find that the percentage fraction difference is well over 15 percent for some of the chosen systems, and is as large as 57 percent in the case of RbMg. We therefore expect that while the error due to missing higher order excitations are usually reasonably small for polarizabilities, the PDM is much more sensitive to this factor. Hence, we defer detailed analysis to a future study and conservatively set an error estimates in the PDM as mentioned in the beginning of this paragraph. 

Another factor that needs to be taken into account in determining the precision of our results is the dependence on the choice of $\epsilon$. We again choose our representative molecules LiBe, KBe and RbBe. We check the change in our results in the neighborhood of $\epsilon = 10^{-4}$ a.u., as shown in Table~\ref{tab5}, and find that the results are stable in that range, with negligible variation. 

\begin{table}[t]
 \centering
\caption{The estimated $\bar\alpha$ values (in a.u.) of the considered Alk-AlkE molecules using the empirical relation given by Eq. (\ref{eqem0}). In the table, the entry in the first row and third column of the results, for example, corresponds to LiCa. }
\begin{tabular}{p{2.0cm}|p{1.4cm}p{1.4cm}p{1.4cm}p{1.4cm}}
\hline \hline 
\diagbox{Alk}{AlkE}& Be & Mg & Ca & Sr \\
\hline \\
 Li & 218.41 & 250.69 & 348.71 & 374.54 \\
 Na & 216.81 & 250.82 & 353.17 & 386.97 \\
 K &  344.42 & 381.07 & 495.26 & 544.84 \\
 Rb & 375.47 & 412.58 & 527.57 & 583.29 \\
 \hline \hline
\end{tabular}
\label{tab6}
\end{table}

\begin{figure}
    \setlength{\tabcolsep}{2mm}
        \begin{tabular}{c}
    \includegraphics[height=6.0cm,width=8.5cm]{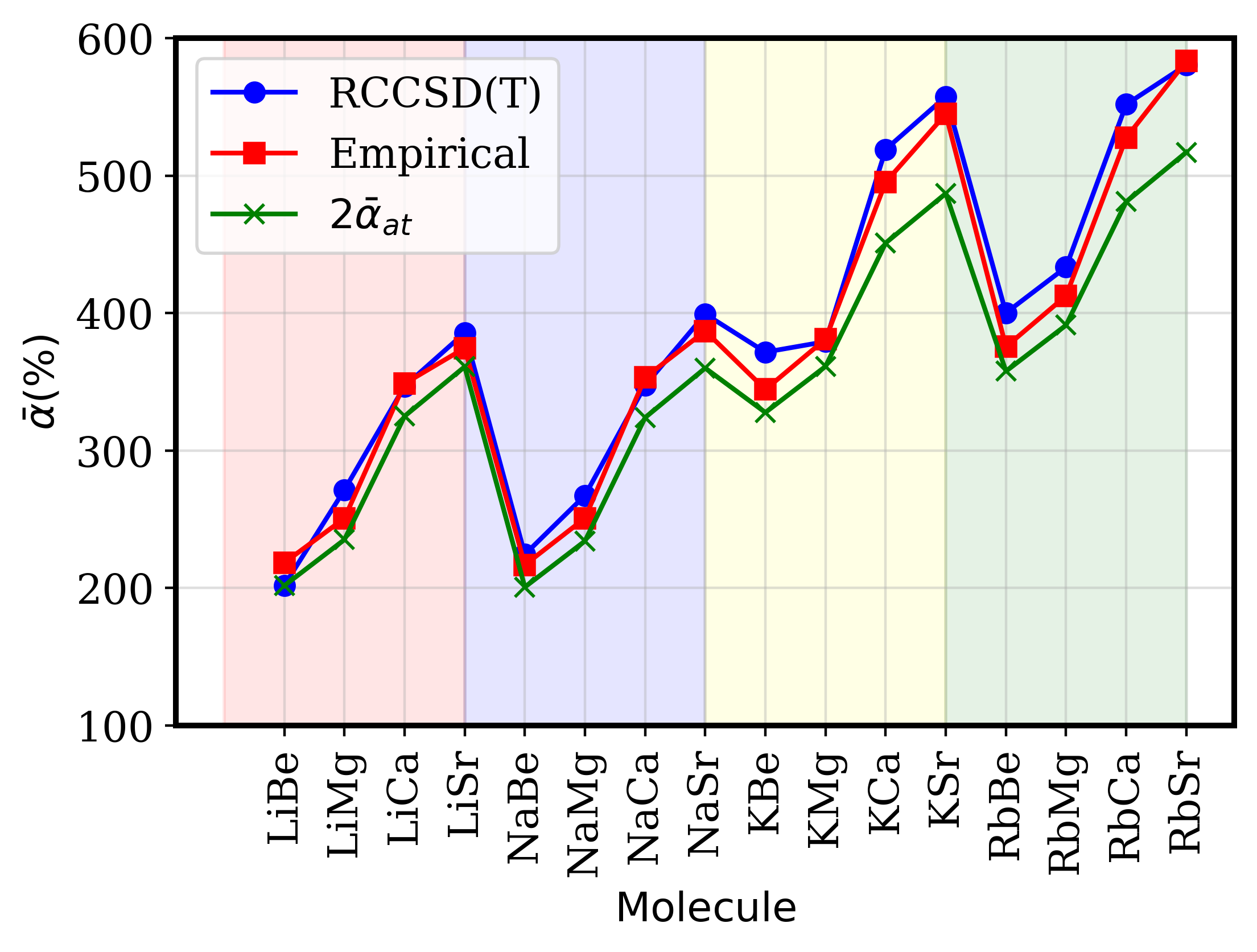} \\ 
        (a) \\
     \includegraphics[height=6.0cm,width=8.5cm]{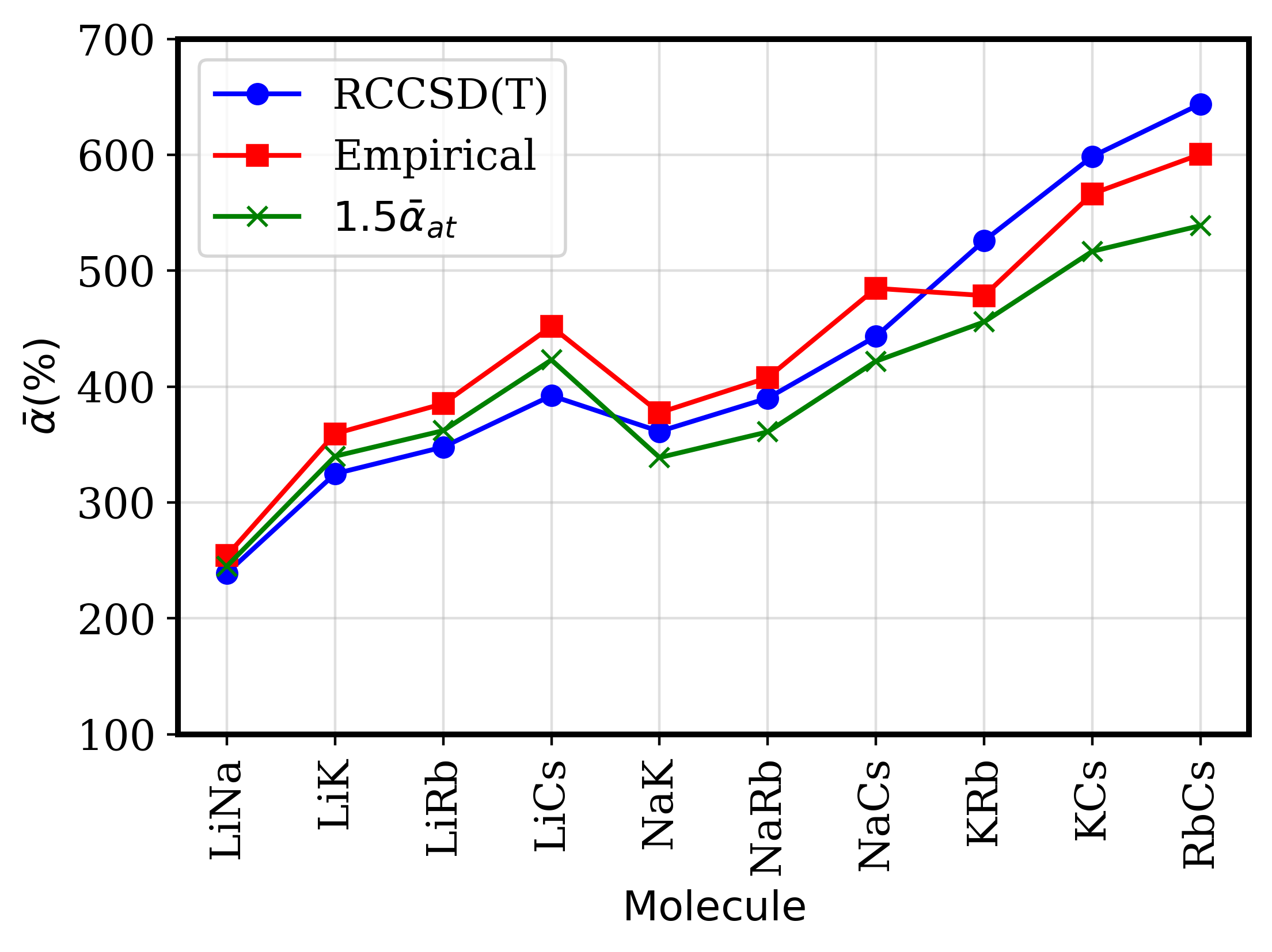}\\   
        (b) \\
    \includegraphics[height=6.0cm,width=8.5cm]{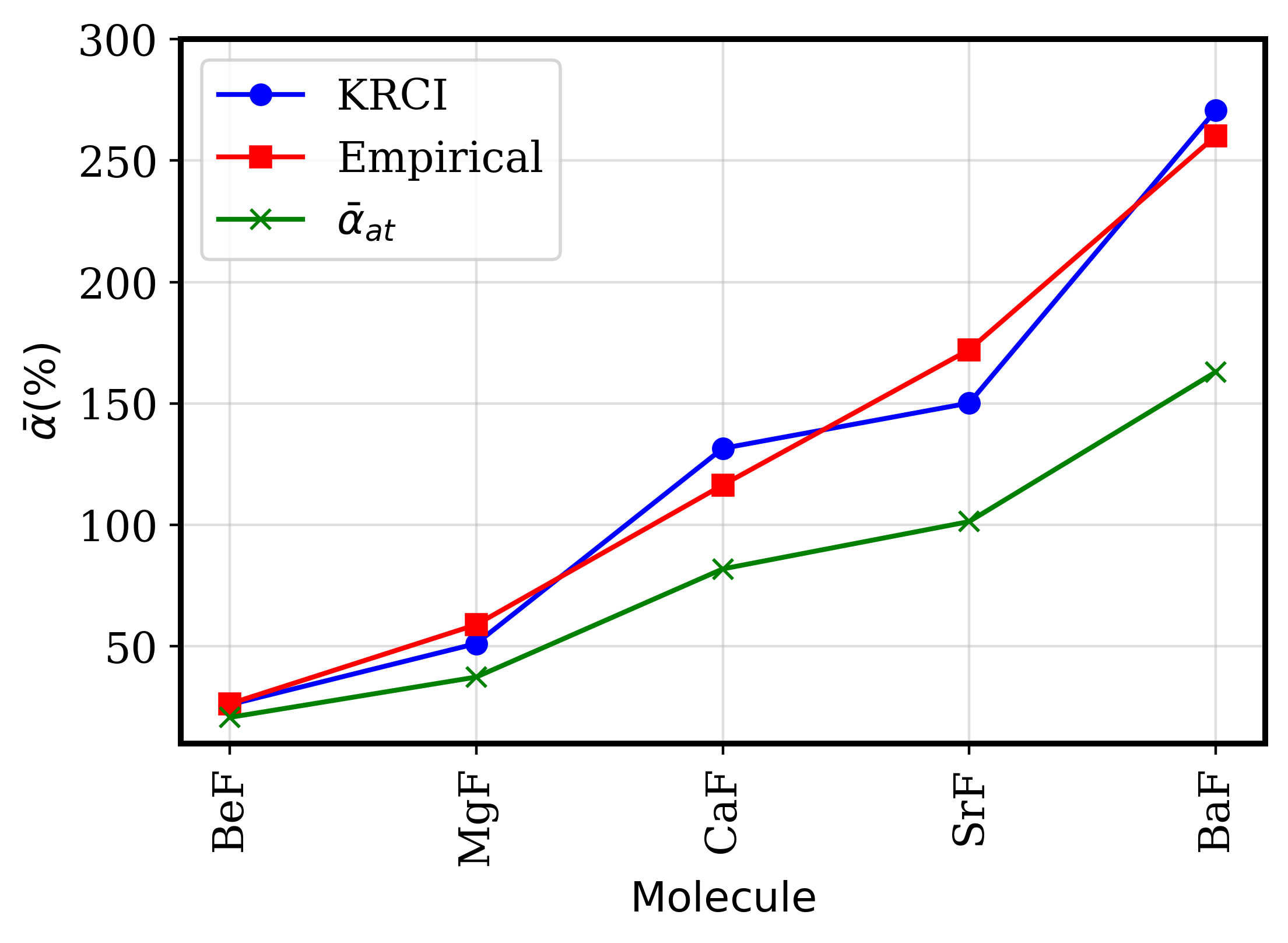}\\
    (c)\\
        \end{tabular}
    \caption{(a) Plot showing the agreement between average polarizabilities calculated using RCCSD(T) method (in blue) and our empirical relation (in red). The dominant part of the empirical relation is shown in green. We find that the red and the blue curves agree to within 10 percent for each of the points, whereas the green curve deviates from the blue one for heavier systems. The shaded background regions have been given to distinguish between families. Sub-figures (b) and (c) show the same plots, but for alkali-alkali molecules and alkaline-earth--monofluorides, respectively. 
    The RCCSD(T) values of PDM and $\bar \alpha$ given in sub-figure (b) were taken from our previous work~\cite{pol1}, whereas the ab initio values of PDM and $\bar \alpha$ in sub-figure (c) were taken from Ref.~\cite{AEMF} and Ref.~\cite{Renu},  respectively. }
\label{empirical}
\end{figure}

\begin{table*}[t]
\centering
\caption{Our final recommended results for $\mu$, $\alpha_\parallel$, $\alpha_\perp$, $\bar\alpha$, and $\Delta\alpha$ from RCCSD(T) calculations, along with the estimated uncertainties that are quoted in the parentheses. We have also compared our results with the previously reported values using MRCI and CCSD(T) methods. $\bar{\alpha}$ and $\Delta \alpha$ are rounded-off to the nearest whole number, given their large values. All the results are given in a.u..}
\begin{tabular}{p{1.4cm}p{1.4cm}p{1.4cm}p{1.2cm}p{1.0cm}p{1.0cm}p{1.7cm}p{1.5cm}}
\hline\hline\\
         Molecule&  $\mu$&$\alpha_\parallel$&$\alpha_\perp$&$\bar\alpha$&$\Delta\alpha$ & Method & Reference  \\
         \hline\\
         LiBe &1.33(18)&376.9&114.12& 202(3)& 263(11) & RCCSD(T) & This work \\
         &1.32&376.55&114.42&201.78&262.13&CCSD(T)&This work\\
         &  1.36&365&& & & MRCI & \cite{Potots} \\
        LiMg &0.40(7)&480.63&166.84&271(6)&314(19) & RCCSD(T)&This work \\
        &0.41&481.64&166.64&271.64&315&CCSD(T)&This work\\
        & 0.46&470&&& & MRCI & \cite{Potots} \\
        LiCa &0.40(6)&584.28&228.08&347(6)&356(26) & RCCSD(T) & This work \\
        &0.43&580.16&229.88&346.64&350.28&CCSD(T)&This work\\
         &0.43&594&230&352&364 & CCSD(T) & \cite{Abe} \\
         &0.47&588&&&& MRCI & \cite{Potots} \\
        LiSr &0.11(2)&620.03&268.5&386(14)&352(44) & RCCSD(T) & This work \\
        &0.16&622.61&276.08&391.59&346.53&CCSD(T)&This work\\
        &0.12&621&271&395&372 & CCSD(T) & \cite{Abe} \\
         &0.11&653&&&& MRCI & \cite{Potots} \\
  & & & & \\
        NaBe &0.86(15)&392.99&140.2& 224(4)& 253(15)& RCCSD(T) & This work\\
        &0.85&390.48&137.79&222.02&252.69&CCSD(T)&This work\\
        &0.92&397&&&& MRCI & \cite{Potots}\\
        NaMg &0.30(7)&437.29&182.39&267(5)&255(11) & RCCSD(T) & This work\\
        &0.31&441.11&183.04&269.06&258.07&CCSD(T)&This work\\
       &0.34&432&&&& MRCI & \cite{Potots}\\
        NaCa &0.43(5)&585.93&242.25&357(5)&344(27) & RCCSD(T) & This work\\
        &0.46&590.94&243.6&359.38&347.34&CCSD(T)&This work\\
        &0.39&581&240&354&361 & CCSD(T) & \cite{Abe}\\
        &0.46&577&&&& MRCI & \cite{Potots} \\
        NaSr &0.20(3)&636.1&280.9&399(7)&355(18) & RCCSD(T) & This work \\
        &0.26&652.97&296.04&415.02&356.93&CCSD(T)&This work\\
        &0.19&633&281&398&352& CCSD(T) & \cite{Abe} \\
        &0.20&636&&&& MRCI & \cite{Potots}\\
   & & & & \\
        KBe &0.75(20)&621.5&246.63&372(13)&375(24)& RCCSD(T) & This work \\
        &0.76&638.91&250.30&379.84&388.61&CCSD(T)&This work\\
        &0.87&628&&&& MRCI & \cite{Potots}\\
        KMg
        &0.35(12)&644.47&291.31&409(12)&353(17) & RCCSD(T) & This work\\
        &0.37&658.97&295.79&416.85&363.18&CCSD(T)&This work\\
        &0.42&656&&&& MRCI & \cite{Potots}\\
        KCa &0.70(13)&888.38&334.4&519(17)&554(36) & RCCSD(T) & This work\\
        &0.76&909.43&330.71&523.62&578.72&CCSD(T)&This work\\
        &0.64&892&326&515&566 & CCSD(T) & \cite{Abe}\\
        &0.83&869&&&& MRCI & \cite{Potots}\\
        KSr &0.51(10)&928.23&372.13&558(18)&556(38) & RCCSD(T) & This work \\
        &0.64&971.74&372.68&572.37&599.06&CCSD(T)&This work\\
        &0.50&942&367&559&574 & CCSD(T) & \cite{Abe}\\
        &0.60&925&&&& MRCI & \cite{Potots}\\
     & & & & \\
        RbBe &0.64(20)&644.08&278.14&400(15)&366(29)& RCCSD(T) & This work\\
        &0.69&692.69&295.05&427.6&397.64&CCSD(T)&This work\\
        &0.78&631&&&& MRCI & \cite{Potots}\\
        RbMg &0.33(13)&667.39&316.44&433(16)&351(19) & RCCSD(T) & This work\\
        &0.37&720.48&335.83&464.05&384.65&CCSD(T)&This work\\
        &0.41&664&&&& MRCI & \cite{Potots}\\
        RbCa &0.70(14)&939.92&357.71&552(20)&582(35)& RCCSD(T) & This work\\
        &0.81&1015.59&368.26&584.04&647.33&CCSD(T)&This work\\
        &0.68&961&357&558&604 & CCSD(T) & \cite{Abe}\\
        &0.86&922&&&& MRCI & \cite{Potots}\\
        RbSr &0.58(11)&970.25&385.59&581(22)&585(41) & RCCSD(T) & This work\\
        &0.72&1080.95&414.26&636.49&666.69&CCSD(T)&This work\\
        &0.55&1009&394&599&615 & CCSD(T) & \cite{Abe}\\
        &0.64&972&&&& MRCI & \cite{Potots}\\
     \hline\\
\end{tabular}
\label{tab7}
\end{table*}
 
\subsection{A simple empirical relation between PDM and average polarizability}  

In this subsection, we seek to explore  interesting connections between $\mu$ and $\bar{\alpha}$. Previous works have sought such empirical relationships, but they usually give complicated functions; for example, the authors in Ref.~\cite{Potots} find such a function connecting PDM and average atomic polarizabilities. We intend to find a relatively simple relation, $\bar{\alpha} = f(\mu, Z_{Alk}, Z_{AlkE},\bar \alpha_{at})$, whose predictions for $\bar{\alpha}$ should agree reasonably well with our RCCSD(T) results. Here, $Z_{Alk}$ and $Z_{AlkE}$ are the atomic numbers of the alkali and alkaline earth atoms, and $\bar \alpha_{at}$ is the average atomic polarizability, given by $ {\bar \alpha}_{at} = (\alpha_{Z_{Alk}} + \alpha_{Z_{AlkE}})/2$. We find that $\bar{\alpha}$ predicted by the empirical relation  
\begin{eqnarray}
\bar \alpha = [a {\bar \alpha}_{at} + \pi ln(Z_{Alk})+b\mu Z_{AlkE}]/c, \label{eqem0} 
\end{eqnarray}
with $a=2$, $b=2.5$, and $c=1$. This agrees reasonably well with our RCCSD(T) results of $\bar{\alpha}$. We note that the dominant contribution comes from the first term, that is, the term containing average atomic polarizability. However, as Fig.~\ref{empirical}(a) shows, using just the first term does not reproduce the RCCSD(T) results for $\bar{\alpha}$ well for heavier systems in each family. The next two terms play an important role in improving the predictability of the empirical relation. Within those two terms, the result is not strongly influenced by the term containing the  atomic number of the alkali atom, owing to the $log$ dependence. An interesting occurrence in the expression is the presence of $\pi$. We have used the ${\bar \alpha}_{at}$ values from Ref. \cite{Schwerdt} in our analyses. In Table~\ref{tab6}, we have listed $\bar\alpha$ estimated using the above relation. By comparing these results with the corresponding values from Table \ref{tab2}, it is clear that the relation predicts $\bar \alpha$ well within 10 percent of the RCCSD(T) values. It is worth commenting at this point that we have made a conscious choice to exclude the dependence of the function on electronegativity differences, as the PDMs that we obtained do not follow the straightforward and simplistic dependence on electronegativity differences. 

We take the applicability of the empirical relation a step further, and check if we can reproduce the correct value of molecular polarizability of LiBa, given its PDM. For this purpose, we use the RCCSD(T) results for the PDM, and find that the average polarizability thus obtained using our empirical relation (459.09 a.u.) matches remarkably well with the RCCSD(T) result for the same quantity (443.69 a.u.), to within 4 percent. 

Lastly, we check if the function from Eq.~(\ref{eqem0}) can faithfully predict $\bar \alpha$s for at least a few other diatomic molecules. For this purpose, we choose two other systems that find important applications, heteronuclear Alk-Alk and AlkE-F molecules. For the former, the empirical relation that agrees reasonably well (within 15 percent) with results from RCCSD calculations (both PDM and $\bar \alpha$ have been taken from Ref.~\cite{pol1}) is with $a=1.5$, $b=2.5$, and $c=1$. i.e.
\begin{eqnarray}
\bar \alpha = 1.5 {\bar \alpha}_{at} + \pi ln(Z_{Alk1})+2.5\mu Z_{Alk2}. \label{eqem1} 
\end{eqnarray}
Here, Alk2 can be Li, Na, K, or Rb. For a given Alk2, Alk1 is either Li, Na, K, or Rb. We also test the quality of the results obtained from the above mentioned equation for homonuclear alkali-alkali molecules. We borrow the values of average polarizabilities from Deiglmayr \textit{et al}~\cite{Deig} for comparison, and find that Eq.~(\ref{eqem1}) gives results that are in reasonable agreement (within 10 percent, except in the case of Li$_2$, which gives 20 percent). 

A similar empirical equation is found to give reasonably good agreement (within 15 percent) with {\it ab initio} theory for AlkE-F molecules when $a=2$, $b=2.5$, and $c=2$. i.e. 
\begin{eqnarray}
\bar \alpha = [2 {\bar \alpha}_{at} + \pi ln(Z_{F})+2.5\mu Z_{AlkE}]/2. \label{eqem2} 
\end{eqnarray}

For this purpose, we chose the PDMs from our previous work~\cite{AEMF}, and $\bar \alpha$s from Ref.~\cite{Renu}. The former work uses RCCSD method, while the latter employs Kramers Restricted Configuration Interaction (KRCI) method. It is worth noting that in spite of the values of $\mu$ and $\bar \alpha$ being taken from different relativistic many-body methods, the empirical data generated from Eq.~(\ref{eqem2}) agrees well with theory. This is not very surprising, as the KRCI results for the PDMs agree well with the corresponding RCCSD results for these systems. The causes behind the seemingly unreasonable effectiveness of the simple functional form based on Eq.~(\ref{eqem0}) are unclear, and future studies on these aspects with more systems may shed light on the rationale behind these observations. Nonetheless, we assume that $\bar \alpha$ of other heavier Alk-AlkE molecules, which are not considered here, will also obey our above suggested empirical relation.

\subsection{Recommended values}

After analysing the reliability of our results and possible uncertainties in their evaluation using the RCCSD(T) method, we would like to quantify now their total uncertainties to the investigated properties. From the above discussions, it is clear that the uncertainties to our calculated values of $\mu$, $\bar\alpha$ and $\Delta\alpha$ come mainly from the dependence on basis functions and missing higher order coupled-cluster excitations. From our CBS extrapolation results in an earlier section, we conservatively assign a maximum uncertainty in the PDMs of about 11$\%$, about 1$\%$ uncertainties to $\bar\alpha$, and about 4$\%$ to $\Delta\alpha$. The uncertainties due to the neglected electron correlation effects are calculated for each of the molecules and are added linearly to the error arising from incompleteness of basis, and are quoted in Table~\ref{tab7} in brackets next to our recommended values, which are our RCCSD(T) results. In the same table, we also give the previously calculated values of some of the above quantities that are obtained using the  MRCI method \cite{Potots} and non-relativistic CCSD(T) method \cite{Abe}. Our RCCSD(T) values and the MRCI values for the PDMs mostly agree within our quoted uncertainties, thus reinforcing our error estimates. It can be noted that the values of $\mu$ from the CCSD(T) calculations in Ref. \cite{Abe} differ from the present CCSD(T) results listed in Table \ref{tab1}. This may be due to the fact that different basis functions (their ANO-RCC contracted bases versus our cc-pCVQZ and Dyall bases) are used in both the works. Moreover, we carry out all-electron calculations, and we expect that to have a non-negligible bearing on the precision of our results, whereas the authors in Ref.~\cite{Abe} freeze inner orbitals. We did not find any other calculations of the $\bar\alpha$ and $\Delta\alpha$ values apart from the CCSD(T) calculations in Ref. \cite{Abe}. 

\section{Conclusion} \label{sec5}

We have investigated the roles of relativistic and electron correlation effects in the determination of the PDMs and static electric dipole polarizabilities of the alkali-alkaline earth heteronuclear dimers. For this purpose, we have employed finite-field coupled-cluster approach in the CCSD(T)) method in both non-relativistic and relativistic frameworks. We find that electron correlation effects can have a significant impact on the final values of the PDMs of some of the considered molecules. Trend-wise, we observe that while correlation effects decrease the PDMs in the Li family, they increase it for the rest of the families, with Mg-containing molecules being exceptions. We also find that for molecules containing Mg, the ratio of the correlation contribution to the PDM to the mean field value of PDM is almost a constant. We find that relativistic effects decrease the PDMs in most cases. The importance of relativistic effects increase from the lighter to the heavier molecules in each family, as expected. Correlation effects play an important role and are pronounced in polarizabilities too, but not as much as in the case of PDMs. Moreover, the observed correlation trends in polarizability anisotropy indicate that the property is likely more sensitive to correlation effects than the average polarizability. Relativistic effects in polarizabilities are found to be non-negligible but are much less prominent than in the case of PDMs. The analysis of relativistic and correlation effects is followed by tests of precision in our results, where we find that the errors due to spin-orbit coupling is negligible. We also test the stability of our numerical results by varying the perturbing parameter in the neighborhood of the chosen value for this work. We observe that the uncertainties due to choice of basis sets and missing coupled-cluster excitations are the dominant error sources. We find a simple and interesting empirical functional form that connects PDM and average polarizability reasonably well and does so consistently, not only for the considered alkali-alkaline earth molecules, but also for homonuclear as well as heteronuclear alkali-alkali systems and for alkaline earth-fluorine molecules. We finally provide  recommended relativistic CCSD(T) values of PDMs and polarizabilities of the alkali-alkaline earth systems and compare them with available literature values as well as our non-relativistic calculations. 

\section*{Acknowledgments}

All the computations were performed on the VIKRAM-100 cluster at Physical Research Laboratory, Ahmedabad. We thank Prof. Trond Saue for providing useful insights on the symmetry aspects of the Dirac program.

\end{document}